\documentclass[10pt,aps,twocolumn,amsmath,amssymb,nofootinbib,superscriptaddress]{revtex4-2}
\usepackage{epsfig} 
\usepackage{amsmath} 
\usepackage{graphicx}
\usepackage{color}
\usepackage{float}
\usepackage{soul,xcolor}
\usepackage[font=scriptsize]{caption}
\usepackage{braket}
\usepackage{bbold}
\usepackage{subfig}
\usepackage{braket}
\usepackage{titlesec}
\usepackage{placeins}
\usepackage{hyperref}
\usepackage{caption}
\begin{document}
\title{NSI effects on tripartite entanglement in neutrino oscillations}

\author{Lekhashri Konwar}
\email{konwar.3@iitj.ac.in }
\affiliation{Indian Institute of Technology Jodhpur, Jodhpur 342030, India}

\author{Bhavna Yadav}
\email{yadav.18@iitj.ac.in}
\affiliation{Indian Institute of Technology Jodhpur, Jodhpur 342030, India}

\begin{abstract}
In this study, we investigate the impact of new physics on different measures of tripartite entanglement within the context of three-flavor neutrino oscillations. These measures encompass concurrence, entanglement of formation, and negativity. We analyze the influence of new physics on these measures across a range of experimental setups involving both reactors and accelerators. Reactor experiments under consideration include Daya Bay, JUNO, and KamLAND setups, while accelerator experiments encompass T2K, MINOS, and DUNE. 
Our analysis reveals that accelerator experiments demonstrate greater sensitivity to NSI, with the most pronounced impact observed in the DUNE experiment. Negativity, while a weaker metric compared to EOF and concurrence, exhibits maximal sensitivity to NSI effects, particularly evident when neutrinos possess moderate to high energies. Conversely, reactor experiments demonstrate less sensitivity to NSI, with concurrence and EOF displaying more prominent effects.

\end{abstract}
\maketitle

\section{Introduction} \label{s:intro}

The phenomenon of neutrino oscillation, validated through a multitude of experimental observations, incontrovertibly signifies the presence of nonzero neutrino masses. This was considered to be the first evidence of physics beyond the standard model (SM), as within SM, neutrino masses were assumed to be zero. The current neutrino oscillation data can be fairly accommodated within the three-flavour framework of standard neutrino oscillations (SO). 

Nonetheless, acknowledging the limitations of the SM as the definitive theory governing fundamental interactions, it becomes imperative to incorporate the potential influence of new physics into the domain of neutrino oscillations. However, there are several beyond SM sub-leading effects that can modify the pattern of neutrino oscillations, for e.g. non-standard interactions (NSI), decoherence, Lorentz invariance violation, and the presence of sterile neutrino. In the current work, we focus on NSI effects.

Numerous ongoing and upcoming experiments are poised to investigate the possible presence of NSI in neutrino oscillations. These experiments range from accelerator-based neutrino oscillation experiments, such as DUNE \cite{DUNE:2015lol} and T2K \cite{T2K:2011qtm}, to reactor experiments like JUNO \cite{JUNO:2015zny} and even astrophysical and cosmological observations involving neutrinos. The quest to uncover NSI effects is driven by the desire to unveil new physics phenomena that could lie hidden within the neutrino sector, ultimately providing vital clues about the fundamental nature of particles and interactions.  The study of quantum correlations in the oscillating neutrino system can also serve as an alternative platform to probe NSI effects.

  Quantum correlations are a multifaceted concept within the domain of quantum mechanics, and they cannot be fully described by a single quantity or measure. Instead, there are various aspects used to characterize these correlations. One of the most famous and intriguing facets of quantum correlations is entanglement. Entanglement serves as a fundamental resource in various quantum information processing tasks, including quantum teleportation \cite{peres1993} and quantum cryptography \cite{ekert}. Importantly, entanglement is not a single-valued quantity; rather, there are numerous methods to quantify it. These quantification approaches include Entanglement of Formation \cite{Bennett,Guo2020}, Concurrence \cite{Wootters1998,Guo2019}, and Negativity \cite{Sabin2008, Vidal2002}.
 
 Entanglement of formation (EOF) is a fundamental concept in the realm of quantum physics. It is defined as the von Neumann entropy of the reduced density matrix, as perceived by any of the three parties involved, namely A, B, or C. In simpler terms, it quantifies the minimum level of entanglement that must be shared among three parties to create a specified quantum state. An intriguing property of EOF is that its value is only zero when the quantum state is separable, underlining its usefulness in identifying entanglement.
 
 Concurrence is another widely used measure for quantifying entanglement. It was introduced as a more tractable alternative to EOF, which involves a minimization step in its definition, making analytical analysis challenging. Concurrence signifies the maximum potential entanglement attainable for a given set of eigenvalues of the density matrix. Its values range between 0 and 1, with 1 indicating maximal entanglement.  Initially, concurrence was defined exclusively for two-qubit states, but recent work by Guo and Gour has extended its applicability to three-qubit states \cite{Guo2019}.
 
Negativity is yet another measure used to characterize entanglement. It is defined as the sum of the absolute values of the negative eigenvalues of the partially transposed density matrix.  One of its significant advantages is its applicability to mixed quantum states, where other measures are more suitable for pure states. Notably, EOF, defined in terms of von Neumann entropy, may not effectively capture entanglement in mixed states since each subsystem possesses non-zero entropy, even in the absence of entanglement. In comparison, negativity provides a more comprehensive characterization of entanglement in mixed states. Therefore, negativity, while a valuable tool, is generally a weaker measure when compared to EOF and concurrence.

Many investigations have employed principles from quantum information theory to yield significant findings in systems of high-energy physics., see for e.g., \cite{Blasone:2007wp,Blasone:2007vw,Banerjee:2014vga,Alok:2014gya,Banerjee:2015mha,Formaggio_2016,Fu:2017hky,Naikoo:2017fos,Naikoo:2018vug,Naikoo:2019eec,Dixit:2019swl,Shafaq:2020sqo,Ming:2020nyc,Blasone:2021mbc,Yadav:2022grk,Blasone:2022iwf,Chattopadhyay:2023xwr,Blasone:2023gau,Caban:2007je,Alok:2015iua,Alok:2024amd,Capolupo:2018hrp,Konwar:2024nwc}. Among these, studies related to quantum correlations in neutrino oscillations are particularly important. 
 However, it is noteworthy that most of these studies operate under the assumption that the dynamics of oscillating neutrino systems adhere to the SO in matter. On the contrary, our study focuses specifically on examining the influence of NSI on different measures of entanglement, such as EOF, concurrence, and negativity. We investigate the effects of NSI across a spectrum of experimental set-ups, spanning various experimental configurations, including reactors and accelerators such as Daya Bay \cite{DayaBay:2012fng, Roskovec:2020rgr}, JUNO \cite{ JUNO:2021vlw}, KamLAND \cite{KamLAND:2002uet}, T2K \cite{ T2K:2013bzi}, MINOS \cite{MINOS:2006foh}, and DUNE \cite{DUNE:2020jqi}, all within the framework of three-flavor neutrino oscillations.
 
 The work is structured as follows: Section \ref{sec2} outlines the formalization of our study in conjunction with matter and NSI. Next, Section \ref{sec3} delves into the discussion of tripartite entanglement measures. Subsequently, Section \ref{sec4} presents our findings, and Section \ref{sec5} encapsulates our concluding remarks.

\section{Formalism} 
\label{sec2}
In the formalism of three-flavor neutrino oscillations, the flavor states $\nu_{e}$, $\nu_{\mu}$, and $\nu_{\tau}$ are expressed as a combination of the mass eigenstates $\nu_{1}$, $\nu_{2}$, and $\nu_{3}$. This relationship is represented by the equation
\begin{equation}\label{1}
    \ket{\nu_{\alpha}}= \sum_{i}U_{\alpha i}\ket{\nu _{i}}, 
\end{equation}
where $\alpha$ represents the flavor indices ($e$, $\mu$, $\tau$), and $i$ represents the mass eigenstate indices (1, 2, 3). This formalism reflects the fact that the initial flavor states at $t=0$ are composed of the underlying mass eigenstates due to the phenomenon of mixing. The $3\times 3$ mixing matrix responsible for this transformation is known as the Pontecorvo-Maki-Nakagawa-Sakata (PMNS) matrix.

The elements of the PMNS matrix are given by
\begin{equation}\label{2}
\begin{pmatrix}
c_{12}c_{13} & s_{12}c_{13}&s_{13}e^{-\iota\delta}\\ 
-s_{12}c_{23}-c_{12}s_{13}s_{23}e^{\iota\delta}& c_{12}c_{23}-s_{12}s_{13}s_{23}e^{\iota\delta }&c_{13}s_{23}\\ 
s_{12}s_{23}-c_{12}s_{13}c_{23}e^{\iota\delta }& -c_{12}s_{23}-s_{12}s_{13}c_{23}e^{\iota\delta}&c_{13}c_{23}
\end{pmatrix}.
\end{equation}
These matrix elements are parameterized by mixing angles, such as $\theta_{12}$, $\theta_{13}$, and $\theta_{23}$, which describe the blending of different flavor components. Here $s_{ij} = \sin \theta _{ij}$ and $c_{ij} = \cos \theta _{ij}$. Additionally, there exists a CP-violating phase, typically denoted as $\delta_{CP}$, which introduces the violation of combined charge conjugation and parity symmetries. While not included in the original matrix, this phase becomes relevant when studying $CP$ violation in neutrino oscillations. In this work, for the sake of simplicity, we will disregard the $CP$-violating phase. 

The determination of the PMNS matrix and its components is an essential aspect of comprehending neutrino oscillations and how neutrinos transition between different flavors during their propagation. Experimental measurements from various neutrino experiments contribute to establishing these parameters.

The evolution of the mass eigenstates is then described as follows
\begin{equation}\label{3a}
     \ket{\nu_{i}(t)}=e^{-\iota  E_{i}t}\ket{\nu_{i }}.
\end{equation}
Here, considering the mass eigenstates at $t=0$ as represented by Eq. \eqref{3a}, the time evolution of the flavor eigenstate can be expressed as follows
\begin{eqnarray}\label{3}
    \ket{\nu _{\alpha}(t)}=\sum_{i}  U_{\alpha i} e^{-\iota  E_{i}t} \ket{\nu_{i}}.
\end{eqnarray}

In the context of the relativistic limit, neutrino flavor states are treated as individual modes. When two modes become entangled, it signifies a form of interconnectedness between their quantum states. Neutrino mode entanglement refers to a specific type of entanglement that emerges among distinct modes within a neutrino field. As a consequence, the time-evolved flavor state can be conceived as an entangled superposition of flavor modes. In the scenario of a three-flavor neutrino system, this concept can be expressed as follows \cite{Blasone:2007vw}
\begin{eqnarray}\label{m4}
    \ket{\nu _{e}}\equiv \ket{1}_{e} \ket{0}_{\mu }\ket{0}_{\tau }\equiv\ket{100}_{e \mu \tau}\notag \\
    \ket{\nu_{\mu }}\equiv\ket{0}_{e } \ket{1}_{\mu } \ket{0}_{\tau }\equiv\ket{010}_{e \mu \tau}\\
   \ket{\nu_{\tau }}\equiv\ket{0}_{e } \ket{0}_{\mu } \ket{1}_{\tau }\equiv\ket{001}_{e \mu \tau}.\notag
\end{eqnarray}
By employing Eq. \eqref{3} and \eqref{m4}, we can derive the time evolution for the tripartite quantum state as follows:
\begin{eqnarray}\label{5}
    \ket{\nu _{\alpha }(t)}=\Bar{U}_{\alpha e}(t)\ket{100}_{e \mu \tau}+\Bar{U}_{\alpha \mu}(t) \ket{010}_{e \mu \tau}\\ \nonumber
    +\Bar{U}_{\alpha \tau}(t) \ket{001}_{e \mu \tau}.
\end{eqnarray}
Here, considering the flavor state $\alpha$ at the initial time $t=0$, and defining $\Bar{U}=U e^{-i \mathcal{H}_m t} U^{\dagger}$ with $\mathcal{H}m = \text{diag}(E_1,E_2,E_3)$ representing the Hamiltonian in the mass eigenbasis having energies $E_{i}$ $(i=1,2,3)$,  Eq. \eqref{5} establishes the entanglement existing at time $t$ among the flavor modes. Consequently, we can utilize the framework of quantum resources to examine the inherent quantum nature within the oscillatory neutrino system.

Next, our attention turns to the density matrix, which will allow us to compute the tripartite entanglement. For the state described in Eq. \eqref{5}, denoted as $\rho _{ABC}^{\alpha}(t) = \ket{\nu _{\alpha }(t)}\bra{\nu _{\alpha }(t)}$, the density matrix is expressed as
\begin{equation}{\label{rho1}}
    \rho _{ABC}^{\alpha}(t)=\begin{pmatrix}
0 & 0 & 0 & 0 & 0 & 0 & 0 & 0\\ 
0 & \rho _{22}^{\alpha } & \rho _{23}^{\alpha } & 0 & \rho _{25}^{\alpha } & 0 & 0 & 0\\ 
0 & \rho _{32}^{\alpha } & \rho _{33}^{\alpha } & 0 & \rho _{35}^{\alpha } & 0 & 0 & 0\\ 
0 & 0 & 0 & 0 & 0 & 0 & 0 & 0\\ 
0 & \rho _{52}^{\alpha } & \rho _{53}^{\alpha } & 0 & \rho _{55}^{\alpha } & 0 & 0 & 0\\ 
0 & 0 & 0 & 0 & 0 &0  & 0 & 0\\ 
0 & 0 & 0 & 0 & 0 &0  & 0 & 0\\
0 & 0 & 0 & 0 & 0 &0  & 0 & 0 
\end{pmatrix},
\end{equation}
where the elements of this matrix can be written as 
\begin{eqnarray}{\label{rhop}}
    \rho _{22}^{\alpha}= \left |\Bar{U}_{\alpha \tau }(t) \right |^{2};~~~~~\notag \\ \rho _{23}^{\alpha}= \Bar{U}_{\alpha \tau }(t)\Bar{U}_{\alpha \mu }^{\ast }(t);\notag \\\rho _{25}^{\alpha}= \Bar{U}_{\alpha \tau }(t)\Bar{U}_{\alpha e }^{\ast }(t);\notag \\
    \rho _{32}^{\alpha}= \Bar{U}_{\alpha \mu }(t)\Bar{U}_{\alpha \tau }^{\ast }(t);\notag \\ \rho _{33}^{\alpha}= \left |\Bar{U}_{\alpha \mu }(t) \right |^{2};~~~~ \\ \rho _{35}^{\alpha}= \Bar{U}_{\alpha \mu }(t)\Bar{U}_{\alpha e }^{\ast }(t); \notag\\
     \rho _{52}^{\alpha}= \Bar{U}_{\alpha e }(t)\Bar{U}_{\alpha \tau }^{\ast }(t);\notag \\ \rho _{53}^{\alpha}= \Bar{U}_{\alpha e }(t)\Bar{U}_{\alpha \mu }^{\ast }(t);\notag \\ \rho _{55}^{\alpha}= \left |\Bar{U}_{\alpha e }(t) \right |^{2}.~~~~~\notag
\end{eqnarray}
The associated probabilities are as follows: $P_{\alpha e}(t)= \left |\Bar{U}_{\alpha e }(t) \right |^{2}$ and $P_{\alpha \mu}(t)= \left |\Bar{U}_{\alpha \mu }(t) \right |^{2}$. In the context of the ultra-relativistic limit, where length is approximately equal to time ($L\equiv t$), and when the energy difference $E_i -E _j$ is considerably smaller compared to the neutrino energy $E$, we can approximate it as $E_i -E j \approx \Delta m_{ij}^{2}/2 E$.

Neutrinos, as they traverse through matter, can engage in interactions through both charged currents (CC) and neutral currents (NC). These interactions can have consequences for the characteristic flavor oscillation pattern. When neutrinos are inside a medium, and these interactions are considered, the Hamiltonian of the SO in matter can be expressed as follows \cite{Giunti:2007ry}
\begin{eqnarray}\label{osc3}
\mathcal{H}_m=\mathcal{H}_{vac}+\mathcal{H}_{mat}~~~~~~~~~~~~~~~~~~~~~~~ ~~~~~\nonumber\\
=\begin{pmatrix}
E_{1} & 0 & 0 \\ 
0 & E_{2} & 0 \\
0 & 0 & E_{3}
\end{pmatrix}+
U^\dagger\begin{pmatrix}
A & 0 & 0 \\ 
0 & 0 & 0 \\
0 & 0 & 0 
\end{pmatrix}U.
\end{eqnarray}
Here, the Hamiltonian for neutrino oscillations in vacuum is denoted as $\mathcal{H}_{vac}$, while $\mathcal{H}_{mat}$ represents the Hamiltonian accounting for oscillations in the presence of matter. Additionally, $A=\pm \sqrt{2}G_{F}N_{e}$ stands for the matter potential, where $G_{F}$ is the Fermi constant and $N_e$ corresponds to the number density of electrons within the matter. Notably, the sign of the matter potential hinges on the type of neutrinos being considered. Specifically, for neutrinos, the matter potential is positive, whereas for antineutrinos, it is negative. In our analysis, the approximate value of the matter potential, denoted as $A$, is around $10^{-13}$ eV, corresponding to Earth's matter density of $\rho= 2.8$ gm/cc.

The impact of NSI is expected to play a subheading role in neutrino flavor oscillations. These NSI effects, affecting both CC and NC interactions, can be incorporated by means of four-fermion dimension-6 operators. It's important to note that constraints on NC-NSI are relatively milder as compared to the highly constrained CC-NSI. In our current analysis, we exclude CC-NSI effects due to these tight constraints. The NC Lagrangian is given by
\begin{eqnarray}\label{nsi1}
\mathcal{L}_{NSI}^{NC}=2\sqrt{2}G_F\sum\limits_{\alpha, \beta, P} \epsilon_{\alpha\beta}^{f,P}(\bar{\nu}_\alpha \gamma^{\mu} P \nu_\beta)(\bar{f} \gamma_\mu P f),
\end{eqnarray}
here $P \in \{P_{R},P_L\}$, $P_{R,L}=(1\mp \gamma^5)/2 $. $P_R$ and $P_L$ are the right and left-handed chirality operators, respectively. $\alpha$ and $\beta$ correspond to different neutrino flavours, $l_{\beta} = e, \mu, \tau$ and $\{f\}\in \{e,u,d \}$.
The strength of NC-NSI is measured by $\epsilon_{\alpha \beta }^{f}$, which is the dimensionless coefficient. It measures the strength as compared to weak interaction coupling constant $G_F$, $i.e.$, $\epsilon_{\alpha\beta}^{f, P}\sim \mathcal{O}(G_x/G_F)$. 

The Hamiltonian in the presence of NSI can be written as
\begin{eqnarray}\label{nsi2}
\mathcal{H}_{tot}&=&
\mathcal{H}_{vac}+\mathcal{H}_{mat}+\mathcal{H}_{NSI}
=
\begin{pmatrix}
E_{1} & 0 & 0 \\ 
0 & E_{2}& 0 \\
0 & 0 & E_{3}

\end{pmatrix}\nonumber\\&&+U^{\dagger } A\begin{pmatrix}
1+\epsilon_{ee}(x) &\epsilon_{e\mu}(x)  &\epsilon_{e\tau}(x) \\ 
\epsilon_{\mu e}(x) & \epsilon_{\mu \mu}(x) & \epsilon_{\mu \tau}(x)\\
 \epsilon_{\tau e}(x) & \epsilon_{\tau \mu}(x) & \epsilon_{\tau \tau}(x)
\end{pmatrix} U.
\end{eqnarray}

The NSI parameters $\epsilon_{\alpha\beta}(x)$, where $\alpha, \beta = e, \mu, \tau$, can be expressed as follows
\begin{equation}\label{nsi3}
    \epsilon _{\alpha \beta }(x)=\sum_{f=e,u,d}\frac{N_{f} (x) }{N_{e}(x)} \epsilon _{\alpha \beta }^{f},
\end{equation}
where $x$ denotes the position or location within the medium, $f$ is summed over the fermions $f \in {e, u, d}$ and $ N_{f} (x) $ is the matter fermion density.  According to the charge neutrality condition ($N_p=N_e$) and from the neutron and proton quark structure we obtain, $N_{u}(x) = 2 N_{p}(x) + N_{n}(x)$ and $N_{d}(x) =  N_{p}(x) + 2 N_{n}(x)$. Substituting these conditions in Eq. \eqref{nsi3}, the expression for $\epsilon_{\alpha\beta}(x)$ turns out to be
\begin{equation}
    \epsilon_{\alpha \beta }(x)=\epsilon_{\alpha \beta }^{e}+(2+Y_{n}(x))\epsilon_{\alpha \beta }^{u}+(1+ 2 Y_{n}(x))\epsilon_{\alpha \beta }^{d},
\end{equation}
where $Y_{n}=N_n(x)/N_e(x)$. NSI parameters can take on both real and complex values. Generally, diagonal NSI parameters have real values, whereas off-diagonal parameters have complex values. In the case of complex NSI parameters, the off-diagonal elements in the flavor basis are generally not equal. From the  Hermiticity condition $\epsilon_{\alpha\beta}=\epsilon_{\beta\alpha}\ast $ i.e. $\epsilon_{\alpha\beta}$ is equal to the complex conjugate of $\epsilon_{\beta\alpha}$. For real NSI parameters, however, the condition $\epsilon_{\alpha\beta}=\epsilon_{\beta\alpha}$ holds, ensuring symmetry. 

Furthermore, NSI can exhibit both axial vector ($A$) and vector ($V$) types, with $\epsilon_{\alpha\beta}^f=\epsilon_{\alpha\beta}^{f,L} \pm \epsilon_{\alpha\beta}^{f,R}$ ('$-$' indicating axial vector and '$+$' indicating vector). These different types of NSI can contribute to neutrino flavor oscillations in distinct ways.

The constraints on NSI parameters are derived from a comprehensive analysis of data from a range of experiments, including both oscillation and non-oscillation studies. This collective analysis aims to determine the allowed values and limitations on NSI parameters. References such as \cite{Esteban:2018ppq, Esteban:2019lfo, Coloma:2019mbs, Coloma2023} are instrumental in providing these constraints based on a global evaluation of experimental data.

In the ultra-relativistic limit, the time evolution operator for a neutrino mass eigenstate is given by $U_{m}(L) = e^{-i \mathcal{H}_{tot}L}$. In the context of three-flavor neutrino oscillations, the evolution operator $U_{m}(L)$ can be expressed as follows \cite{Ohlsson:1999xb}
\begin{eqnarray}\label{osc4}
U_{m}(L)&=&e^{-i\mathcal{H}_{tot}L}=\phi ~e^{-iLT}\\
&=&\phi \sum_{a=1}^{3}e^{-i L\lambda _{a}}\frac{1}{3\lambda _{a}^2+c_{1}}\left [( \lambda _{a}^{2}+c_{1}) I+\lambda _{a}T+T^{2}\right ].\nonumber
\end{eqnarray}
 
 Here $\phi$ is a complex phase factor, which is defined as $ \phi = e^{-\iota L (tr H_{tot})/3}$, where $L$ is the length between source and detector, while $H_{tot}$ is total Hamiltonian which accounts for vacuum, matter and NSI effects. {\color{black}The coefficient $c_{1}$ is real, which is defined as the multiplication of the determinant of T and trace of $ T^{-1} $.} $\lambda _{1}$, $\lambda _{2}$ and $\lambda _{3}$ are the eigenvalues of the matrix $T$, which is given as
\begin{equation}\label{osc5}
T \equiv \mathcal{H}_{tot}- (tr H_{tot})I/3 =
\begin{pmatrix}
T_{11} & T_{12}& T_{13} \\
T_{21} & T_{22}& T_{23} \\
T_{31} & T_{32}& T_{33} 
\end{pmatrix}.
\end{equation}
The evolution operator $U_{f}(L)$, which operates in the flavor basis, can be computed as $U_{f}(L)=U^{\dagger} U_{m}(L) U$. This evolution operator in the flavor basis is essential for calculating the probabilities of neutrino oscillations. To calculate the probabilities, we focus on the matrix elements of the evolution operator in the flavor basis. By considering the absolute values of these elements and squaring them, we obtain the oscillation probabilities for the transition from $\nu_{\alpha}$ to $\nu_{\beta}$. These probabilities are expressed as follows
\begin{equation}
   \Tilde{ P}_{\alpha \beta }\equiv \left|A_{\alpha \beta } \right|^{2}= \left| \bra\beta U_{f}(L) \ket\alpha\right|^{2}.
\end{equation}

In this context, $\Tilde{ P}_{\alpha \beta }$ represents the probability of neutrino oscillation in the presence of NSI. In the formalism described above, the mass evolution operator $U_{m}(L)$ depends on the Hamiltonian of the system. When considering vacuum oscillations, the operator $U_{m}(L)$ is constructed using the vacuum Hamiltonian $\mathcal{H}_{vac}$. On the other hand, when neutrinos interact with matter, the operator $U_{m}(L)$ involves the matter Hamiltonian $\mathcal{H}_{m}$, which accounts for both vacuum and matter effects. In the presence of NSI, the operator $U_{m}(L)$ is based on the total Hamiltonian $\mathcal{H}_{tot}$, which encompasses contributions from vacuum, matter, and Non-Standard Interactions. The choice of the appropriate Hamiltonian depends on the specific context being considered, whether it's vacuum, matter, or the inclusion of NSI effects.

In the upcoming section, we delve into the description of tripartite entanglement measures, both in terms of probabilities and through the lens of reduced density matrices. It's evident that these measures can be computed by leveraging the formalism elucidated in the preceding section.

\section{ Tripartite Entanglement Measures }\label{sec3}
In the tripartite entanglement scenario, the quantum states of three particles are intertwined in a manner where the state of any one particle cannot be explained separately from the states of the other two particles. This implies that measurements conducted on one particle can immediately affect the states of the other two particles, irrespective of the spatial separation between them. In the realm of neutrino oscillations, the notion of tripartite entanglement emerges from blending the three flavor eigenstates with the underlying mass eigenstates. This combination results in quantum superposition in which the state of one flavor becomes intricately connected with the others, surpassing their distinct identities. After suitable approximations,  we can also consider a two-flavor neutrino oscillation scenario that can allow us to quantify entanglement in terms of bipartite quantum correlation measures involving two parties. These bipartite measures provide a concise framework for understanding the correlation between two flavor states \cite{Jha, jha2}. However, in this work, we focus on tripartite entanglement.
 
 In this section, we will provide a succinct overview of some of the tripartite entanglement quantum correlation measures that have been employed in this study.

\emph {Entanglement of Formation}: Two quantum systems, A and B, are represented by a density matrix. We can decompose this density matrix into all possible pure-state ensembles $\ket{\psi_i}$ and their associated probabilities $p_i$. Mathematically, this can be expressed as
\begin{equation}
\rho = \sum_i p_i \ket{\psi_i}\bra{\psi_i},
\end{equation}
 where $\rho$ denotes the density matrix for composite systems A and B, $\ket{\psi_i}$ denotes the pure states in the ensemble, and $p_i$ denotes the corresponding probabilities. The completeness of the ensemble is ensured by the requirement that the sum of the probabilities over all pure states equal to one. This decomposition facilitates the representation of the density matrix in terms of its constituent pure states and their probabilities, thereby providing a more detailed understanding of the quantum correlations inherent in the system.

To quantify the entanglement within subsystems A or B, the entanglement of formation (EOF) is defined for each pure state within the ensemble as:
\begin{equation}
    EOF(\ket{\psi_{i}})=-Tr(\rho _{A} \log_{2}\rho _{A})=-Tr(\rho _{B} \log_{2}\rho _{B}),
\end{equation}
here $\rho_{A(B)}$ represents the partial trace of the density matrix $\rho$, yielding the reduced density matrix of subsystem A (B). The entanglement characterized by the EOF can then be generalized to encompass an arbitrary tripartite pure state $\rho_{ABC}(t)$. In this context, the EOF for such a tripartite state is often defined as follows \cite{Guo2020}
\begin{equation}
   EOF(\rho _{ABC}(t))=\frac{1}{2}[S(\rho _{A})+S(\rho _{B})+S(\rho _{C})],
\end{equation}
where $\rho_{A}$, $\rho_{B}$ and $\rho_{C}$ are reduced density matrices which are $\rho _{A}=Tr_{BC}(\rho _{ABC}(t))$, $\rho _{B}=Tr_{AC}(\rho _{ABC}(t))$ and $\rho _{C}=Tr_{AB}(\rho _{ABC}(t))$. Further $ S_({\rho _{A}})$, $ S_({\rho _{B}})$ and $ S_({\rho _{C}})$ are von Neumann entropies defined as $S_({\rho _{A}})=-Tr(\rho _{A}\log\rho _{A})$ and same with $S_({\rho _{B}})$ and $S_({\rho _{C}})$.

EOF in terms of survival and oscillation probabilities for vacuum can be presented as follows \cite{Ming2021}:
\begin{eqnarray}\label{EOF}
    EOF^{\alpha}=-\frac{1}{2}[P_{\alpha e} \log_{2}P_{\alpha e}+P_{\alpha \mu} \log_{2}P_{\alpha \mu}+P_{\alpha \tau} \log_{2}P_{\alpha \tau}\nonumber\\
    +(P_{\alpha \mu}+P_{\alpha \tau}) \log_{2}(P_{\alpha \mu}+P_{\alpha \tau})\nonumber\\
    +(P_{\alpha e}+P_{\alpha \tau}) \log_{2}(P_{\alpha e}+P_{\alpha \tau})\nonumber\\
    +(P_{\alpha \mu}+P_{\alpha e}) \log_{2}(P_{\alpha \mu}+P_{\alpha e})].\nonumber\\
\end{eqnarray}
Here $\alpha = e, \mu, \tau$, representing the initial flavor state of the neutrino.

\emph {Concurrence}: Originally introduced for two-qubit states by Wootters in 1998 \cite{Wootters1998}, the concurrence has traditionally been a measure of entanglement limited to such systems. However, a recent study has demonstrated that the concept of concurrence can be extended to quantify entanglement in three-qubit states as well. This expansion of the concurrence's applicability represents a significant advancement in our understanding of entanglement in more complex quantum systems. \cite{Guo2019}
\begin{equation}
    C(\rho _{ABC})=[3-Tr(\rho _{A})^{2}-Tr(\rho _{B})^{2}-Tr(\rho _{C})^{2}]^{\frac{1}{2}},
\end{equation}
where  $\rho _{A}=Tr_{BC}(\rho _{ABC}(t))$, $\rho _{B}=Tr_{AC}(\rho _{ABC}(t))$ and $\rho _{C}=Tr_{AB}(\rho _{ABC}(t))$.

For oscillating neutrinos, the concurrence for vacuum can be expressed in terms of oscillation and survival probabilities as \cite{Ming2021}
\begin{equation}\label{C} 
 C^{\alpha}=\sqrt{3-3P_{S}-2P_{\alpha \mu}P_{\alpha \tau}-2P_{\alpha e}(P_{\alpha \mu}+P_{\alpha \tau})},
\end{equation}
where $P_{S}$ is the sum of the square of probabilities, and it is defined as $(P_{\alpha e}^{2}+P_{\alpha \mu}^{2}+P_{\alpha \tau}^{2})$.

\emph {Negativity}:  In an alternative approach to quantifying entanglement within a system, the concept of negativity (N) is employed \cite{Sabin2008,Vidal2002}. It is defined as
\begin{equation}
    N=(N_{A-BC}N_{B-CA}N_{C-AB})^{\frac{1}{3}},
\end{equation}
where $N_{A-BC}=-\sum_{i}\lambda _{i}^{A}$, $N_{B-CA}=-\sum_{j}\lambda _{j}^{B}$ and $N_{C-AB}=-\sum_{k}\lambda _{k}^{C}$. Here $\lambda _{i}^{A}$, $\lambda _{j}^{B}$ and $\lambda _{k}^{C}$ are negative eigenvalues of $\rho _{ABC}^{T_{\alpha }}(t)$, which is partial transpose of matrix $\rho_{ABC}(t)$.

In terms of survival and oscillation probabilities, negativity is given as \cite{Ming2021}:
\begin{equation}\label{N}
    N^{\alpha}=[\sqrt{P_{\alpha e}}\sqrt{P_{\alpha \mu}+P_{\alpha \tau}}\sqrt{P_{\alpha e}}\sqrt{P_{\alpha \mu}}\sqrt{P_{\alpha e}+P_{\alpha \mu}}\sqrt{P_{\alpha \tau}}]^{\frac{1}{3}}.
\end{equation}

\section{Results and Discussion}\label{sec4}

\begin{table}[t]
	\centering
	\caption{Standard neutrino oscillation parameters \cite{Salas} }
	\label{Tab1}
	\begin{tabular}{|c|c|}
		\hline
		Parameters  &  Best fit$\pm 1\sigma$\\
		\hline\hline
		$\theta_{12}^{o}$    &   $34.3\pm 1.0$\\
		\hline
		$\theta_{13}^{o}$    &   $8.58_{-0.15}^{+0.11}$\\
		\hline
		$\theta_{23}^{o}$    &   $48.79_{-1.25}^{+0.93}$\\
		\hline
		$\Delta m_{21}^{2}\times 10^{-5}\, \rm (eV^{2})$  &  $7.5_{-0.20}^{+0.22}$ \\
		\hline
		$\Delta m_{31}^{2}\times 10^{-3}\, \rm (eV^{2})$  &  $2.56_{-0.04}^{+0.03}$ \\
		\hline	
	\end{tabular}
\end{table}

\begin{table}[t]
	\centering
	\caption{NSI parameters \cite{Coloma2023} are given here, and these parameters are assumed to be real. The NSI upper limits have been used to generate the NSI plots.} 
	\label{Tab2}
	\begin{tabular}{|c|c|c|}
		\hline
        NSI Parameters  & Range ($1 \sigma$)& Range ($2 \sigma$)\\      
		\hline\hline
				$\epsilon_{ee}^{\oplus}$ & [-0.30, 0.20] $\oplus$ [0.95, 1.3] & [-1.00, 1.4]\\
		\hline
		$\epsilon_{e \mu}^{\oplus}$ &[-0.12, 0.011]& [-0.20, 0.09] \\
		\hline
		$\epsilon_{e \tau}^{\oplus}$ & [-0.16, 0.083]& [-0.24, 0.30]\\
		\hline
		$\epsilon_{\mu \mu}^{\oplus}$ & [-0.43, 0.14] $\oplus$ [0.91, 1.3]& [-0.80, 1.4] \\
		\hline
		$\epsilon_{\mu \tau}^{\oplus}$ & [-0.047, 0.012]& [-0.021,0.021] \\
		\hline
		$\epsilon_{\tau \tau}^{\oplus}$ &[-0.43, 0.21] $\oplus$ [0.83, 1.3]& [-0.85, 1.4]\\
		\hline
	\end{tabular}
\end{table}

In this section, we delve into an exploration of how NSI affects tripartite entanglement measures within a three-flavor neutrino system. We focus on analyzing prominent entanglement measures, including entanglement of formation, concurrence, and negativity. Our examination encompasses various experimental setups, specifically those involving reactors and accelerators. Through detailed analysis, we aim to discern the impact of NSI on the entanglement characteristics of neutrinos, shedding light on the intricate interplay between quantum correlations and particle interactions. We consider the following reactor and accelerator experimental setups:
\begin{itemize}
    \item Daya Bay ($L \approx2$ km, $E\approx0.8-6$ MeV) \cite{DayaBay:2012fng,Roskovec:2020rgr},
    \item JUNO ($L \approx53$ km, $E\approx1-8$ MeV) \cite{JUNO:2015zny,JUNO:2021vlw},
    \item  KamLAND ($L \approx180$ km, $E\approx1-16$ MeV) \cite{KamLAND:2002uet},
    \item  T2K ($L \approx295$ km, $E\approx0-6$ GeV) \cite{T2K:2011qtm,T2K:2013bzi},
    \item  MINOS ($L\approx735$ km, $E\approx1-10$ GeV) \cite{MINOS:2006foh}, 
    \item DUNE ($L \approx1300$ km, $E\approx1-14$  GeV) \cite{DUNE:2015lol,DUNE:2020jqi}. 
\end{itemize} 

\begin{figure*}[htb]
\includegraphics[scale=0.32]{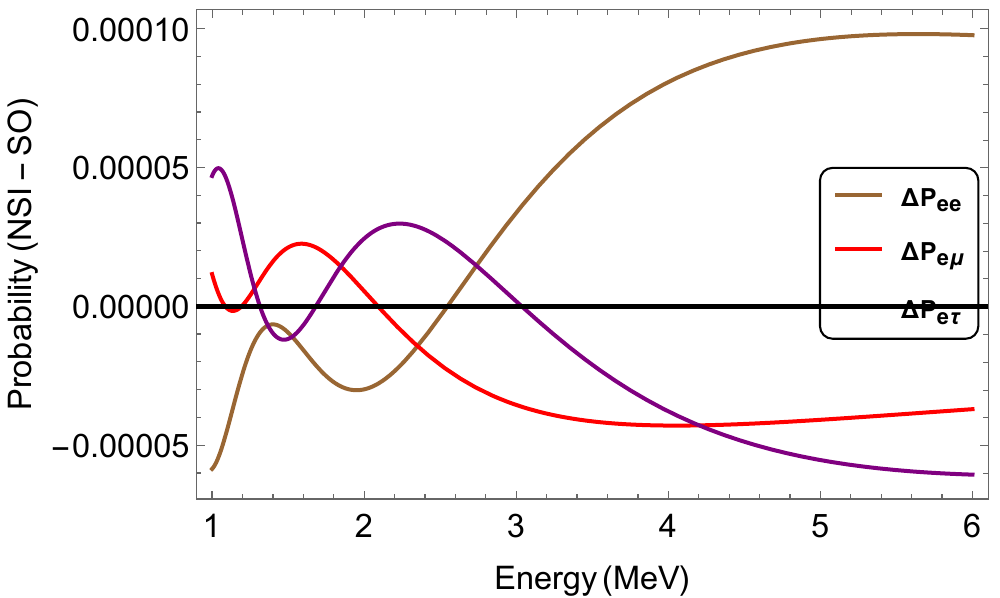}
\includegraphics[scale=0.32]{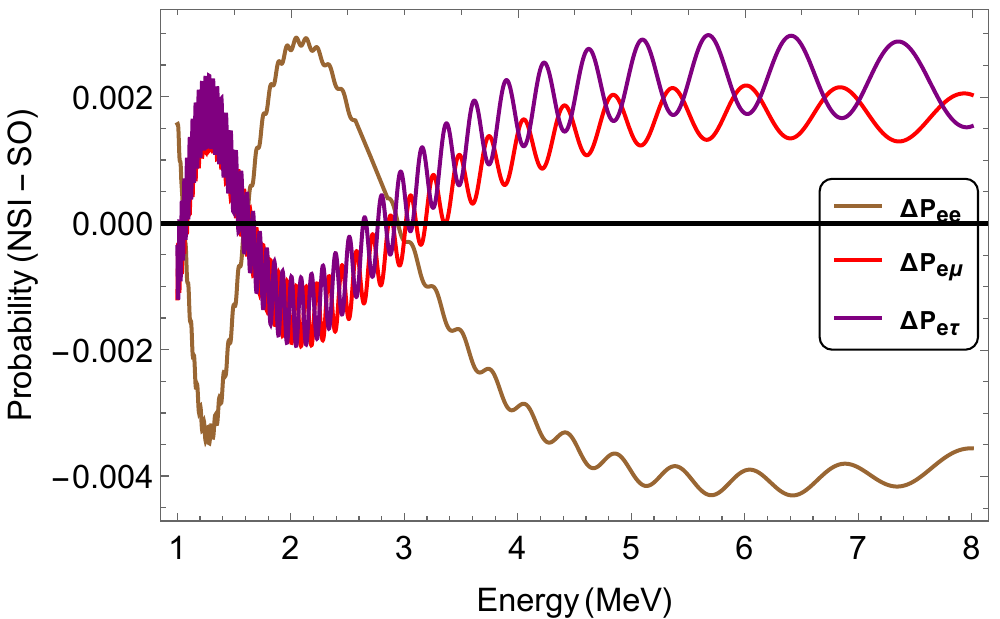}
\includegraphics[scale=0.32]{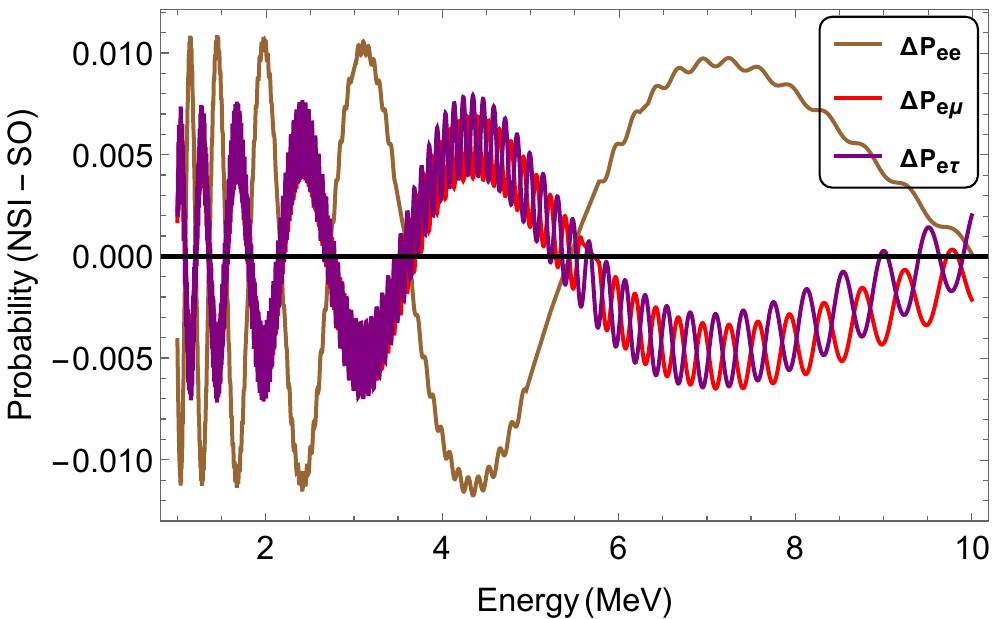}
\\
\includegraphics[scale=0.32]{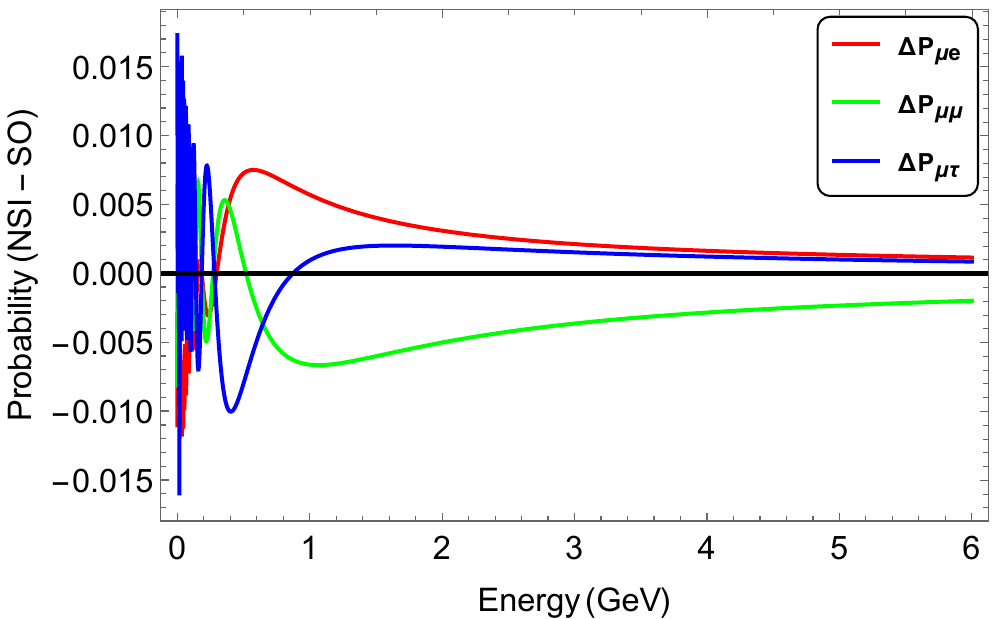}
\includegraphics[scale=0.32]{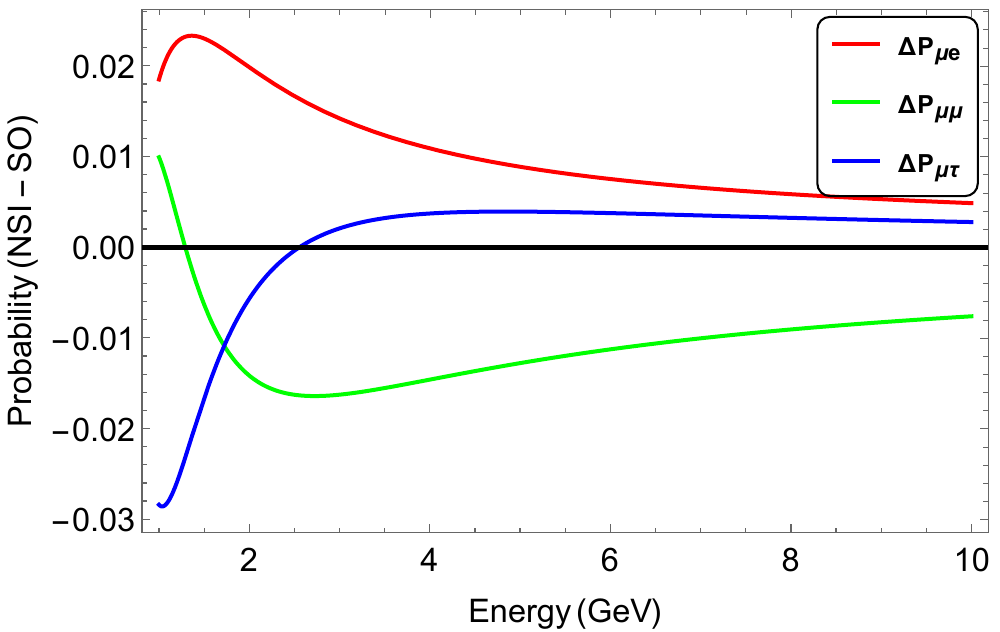}
\includegraphics[scale=0.32]{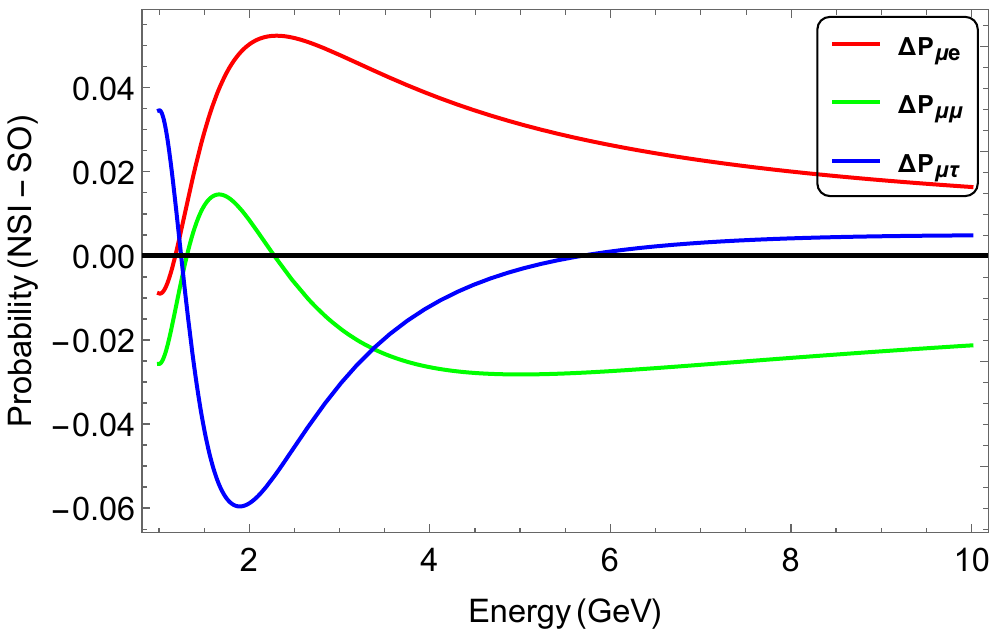}
\caption{Showing variation of probabilities for (NSI$-$SO) scenario with neutrino energies for reactor experiments in upper panel i.e. Daya Bay (left), JUNO (center), KamLAND (right) and for accelerator in lower panel i.e. T2K (left), MINOS (center), DUNE (right). Colour representation is brown for $\Delta P_{e e}$ , red for $\Delta P_{e \mu}$ or $\Delta P_{\mu e}$, purple for $\Delta P_{e \tau}$, green for $\Delta P_{\mu \mu}$ and blue for $\Delta P_{\mu \tau}$. Standard neutrino oscillation parameters and NSI parameters are given in table \ref{Tab1} and table \ref{Tab2}, respectively.}
\label{fig3}
\end{figure*}

The standard neutrino oscillation parameters are presented in Table \ref{Tab1}, while the NSI parameters are provided in Table \ref{Tab2}. These tables pertain to the framework of three-flavor neutrino oscillations. In the context of reactor experiments, the initial neutrino is considered to be the antineutrino $\bar{\nu_e}$, whereas for accelerator experiments, the initial neutrino is denoted as $\nu_{\mu}$.

Figure \ref{fig3} depicts the probabilities for NSI relative to the SO matter effect for the experimental configurations. As all the entanglement measures are expressed in terms of probabilities, we studied the effect of NSI on probabilities. In reactor experiments, where the initial neutrino state is $\nu_{e}$, the probabilities $P_{e e}$, $P_{e \mu}$, and $P_{e \tau}$ are plotted. Conversely, in accelerator experiments, where $\nu_{\mu}$ is the initial neutrino state, the probabilities $P_{\mu e}$, $P_{\mu \mu}$, and $P_{\mu \tau}$ are illustrated. The baseline of the experiments amplifies their sensitivity to NSI effects; a longer baseline results in a more pronounced influence on the probabilities. In reactor experiments, when $P_{e e}$ is enhanced due to NSI, it leads to a reduction in $P_{e \mu}$, $P_{e \tau}$, and vice versa. Overall, it is evident that the effect of NSI on probabilities is negligible in reactor experiments, so the same is expected in quantum correlation measures. In accelerator experiments, $P_{\mu e}$ exhibits slightly more dominance compared to $P_{\mu \mu}$ and $P_{\mu \tau}$.


\begin{figure*}[htb]
\includegraphics[scale=0.40]{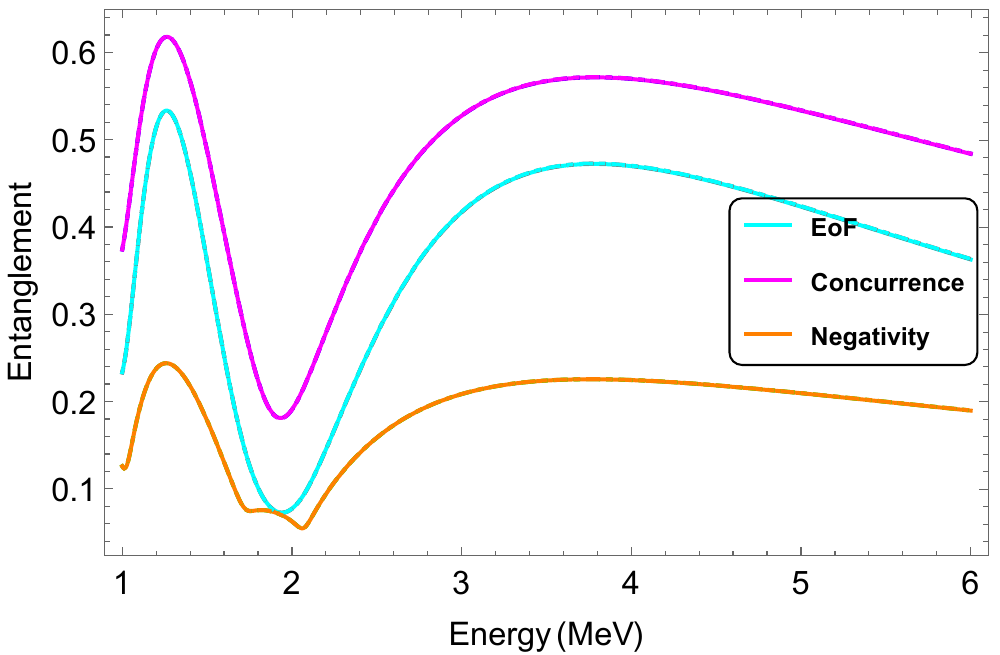}\hspace{1 cm}
\includegraphics[scale=0.40]{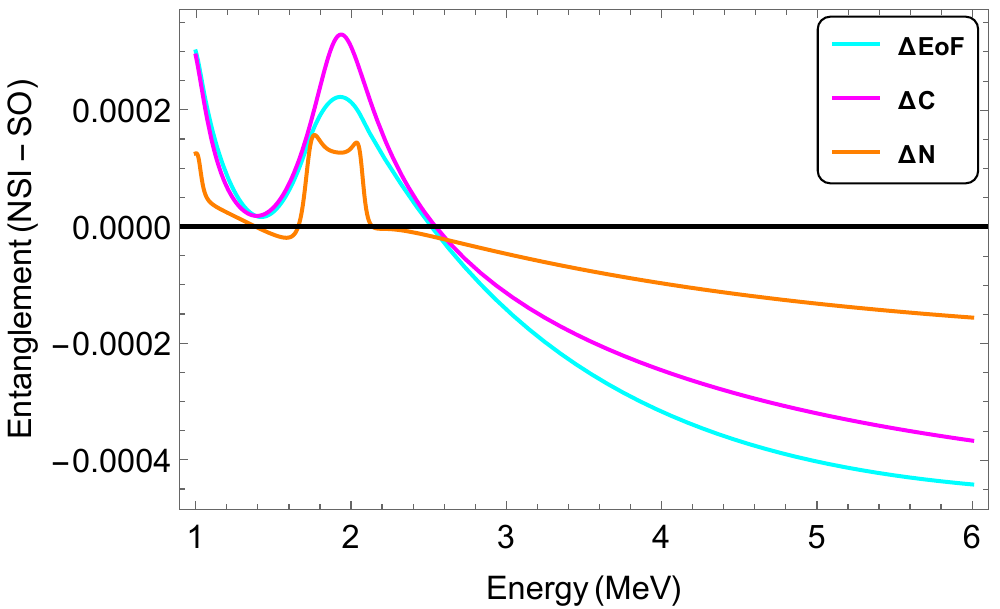}
\\
\includegraphics[scale=0.40]{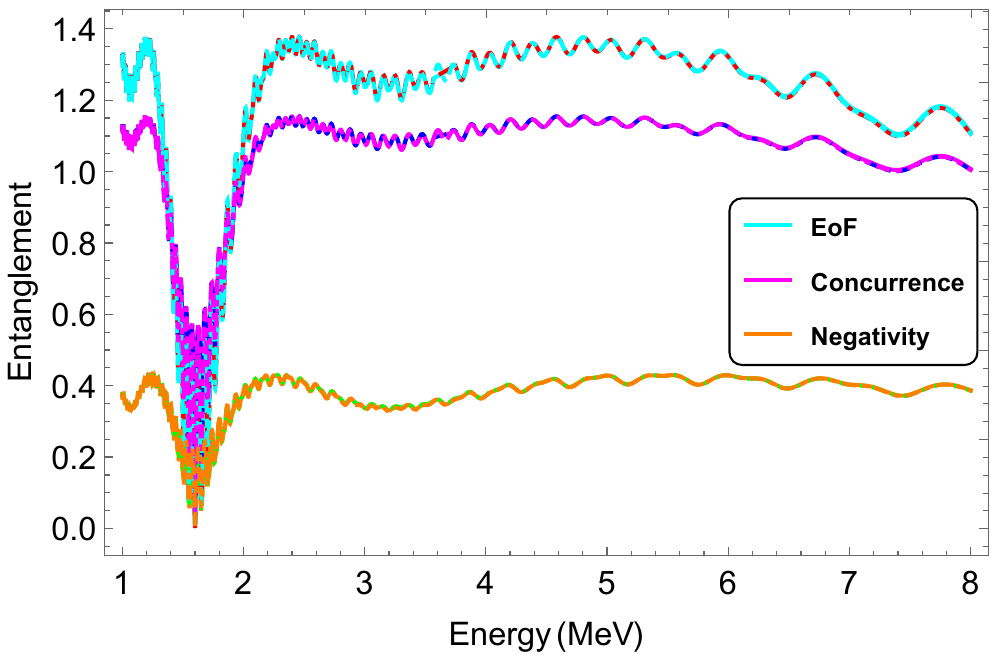}\hspace{1 cm}
\includegraphics[scale=0.40]{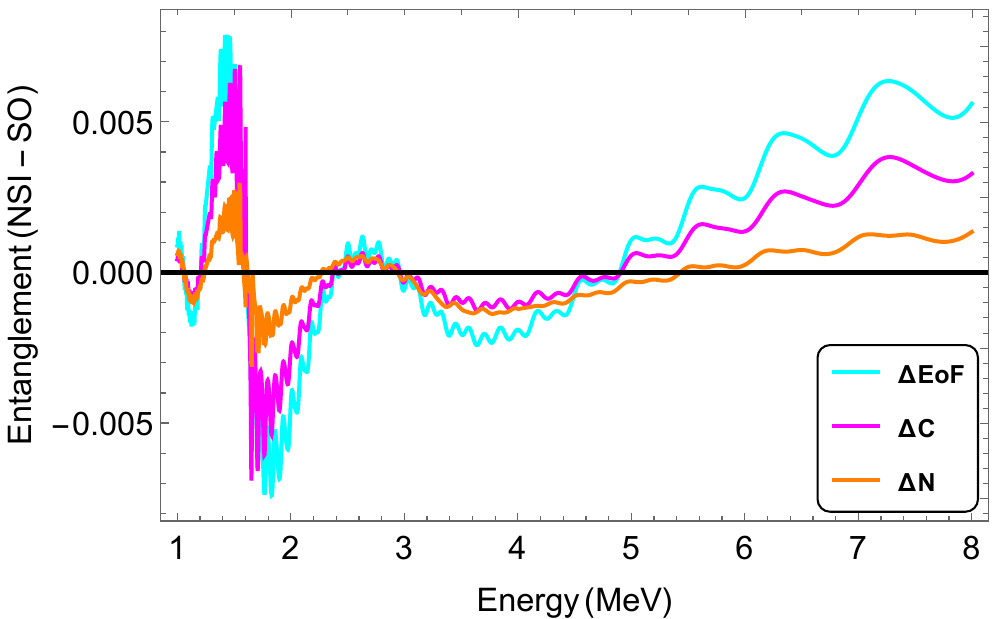}
\\
\includegraphics[scale=0.40]{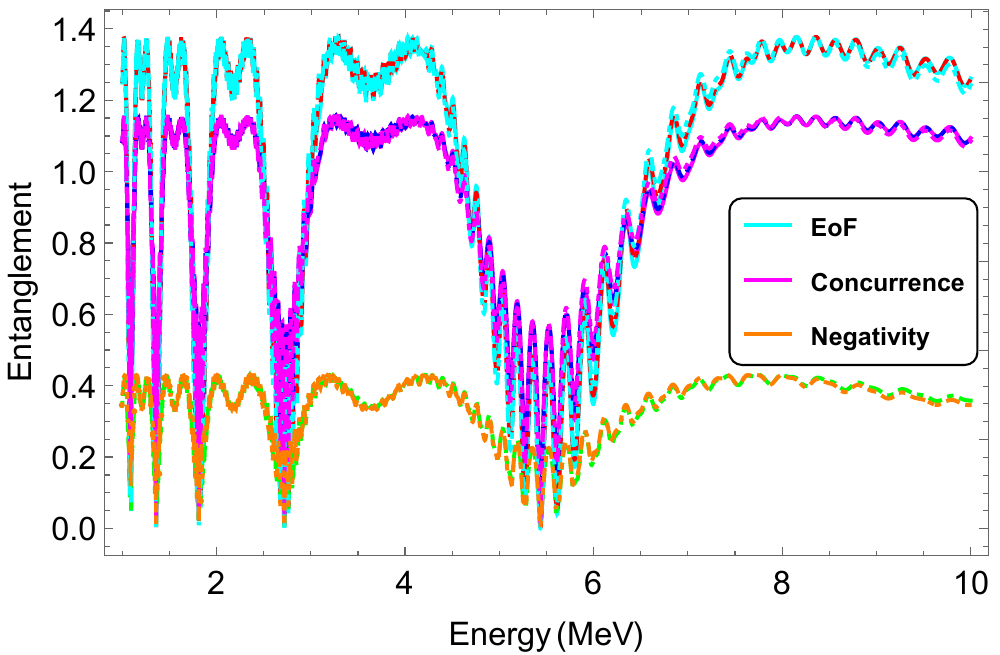}\hspace{1 cm}
\includegraphics[scale=0.40]{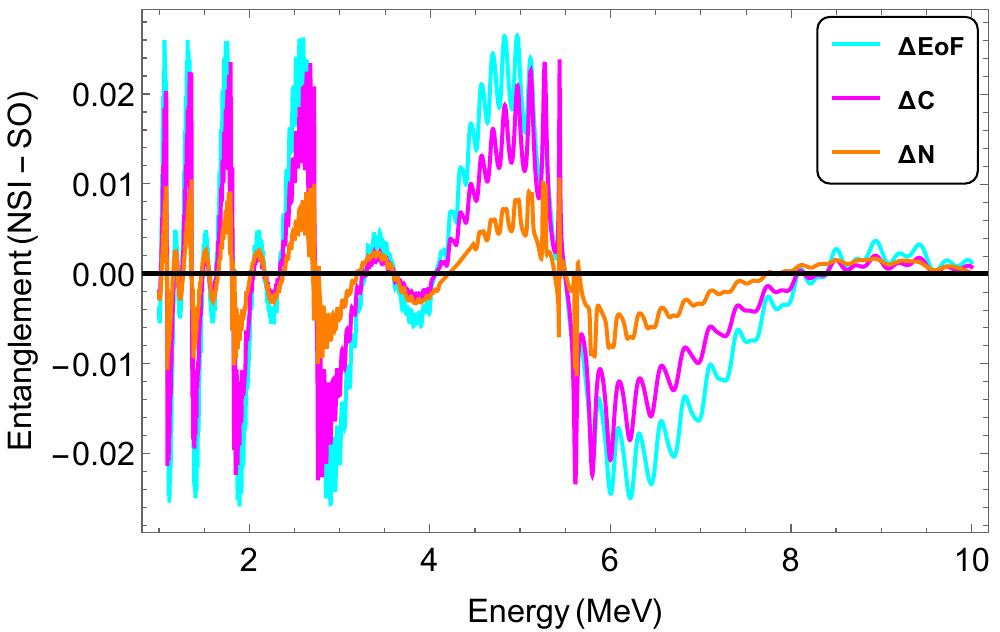}
\caption{The three panels depict the prediction of EOF, concurrence, and negativity for various reactor neutrino oscillation experimental setups, i.e., Daya Bay (upper), JUNO (middle), and KamLAND (lower). Oscillation and NSI parameters are the same as given in figure \ref{fig3}. Colour representation follows cyan for EoF, magenta for concurrence, and orange for negativity. In the left panel, NSI is represented by solid lines, and the vacuum is represented by dot-dashed lines. {\color{black}The SO matter effect is represented by dot-dashed lines, colour representation follows red for EoF, blue for concurrence, and green for negativity. }  }
\label{fig1}
\end{figure*}

Figure \ref{fig1} illustrates the projected results for EoF, concurrence, and negativity in the context of different reactor neutrino oscillation experiments, namely Daya Bay, JUNO, and KamLAND. The right panel of each experiment showcases the differences in these measures, denoted as $\Delta EoF$, $\Delta C$, and $\Delta N$, between the scenario with NSI and the SO in matter scenario. This visual representation serves to emphasize the discernible distinctions between the two scenarios, facilitating a clearer understanding of the impact of NSI on entanglement measures.

The left panel of the diagram clearly showcases a notable trend across all three assessed setups. For the JUNO and KamLAND configurations, the EoF stands out as a notably robust measure of entanglement. In contrast, for the Daya Bay setup, the concurrence exhibits higher values throughout the entire energy range under scrutiny. Conversely, negativity consistently demonstrates weaker values within this energy range.
Significantly, this pattern remains consistent for both the matter SO scenario and the introduction of NSI. This finding underscores the enduring importance of both EoF and concurrence as potent indicators of entanglement, particularly in the context of neutrino oscillation experiments.

\begin{figure*}[htb]
\includegraphics[scale=0.40]{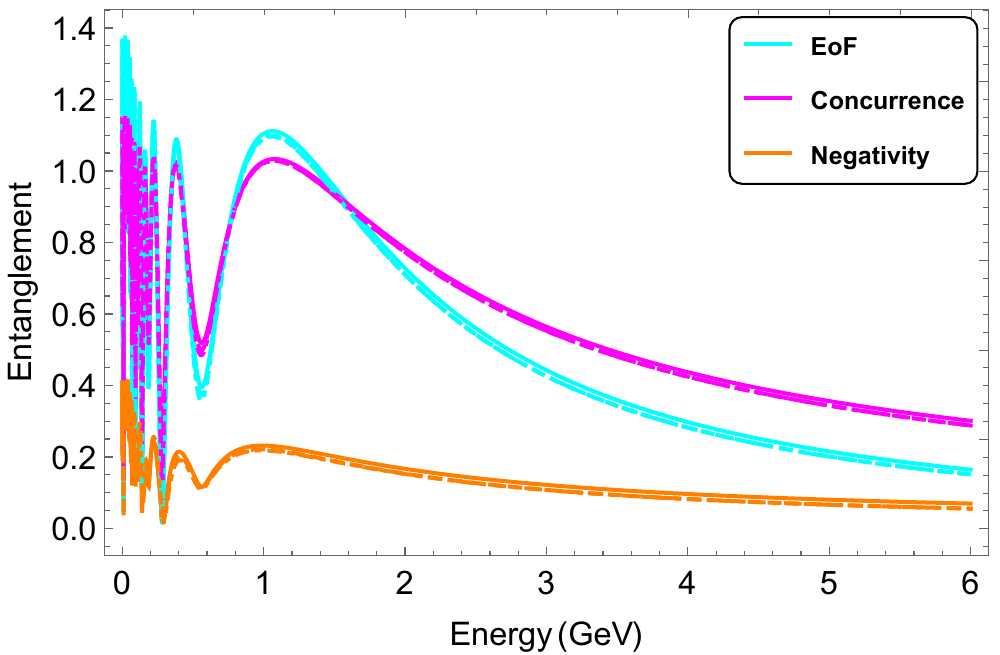}\hspace{1 cm}
\includegraphics[scale=0.40]{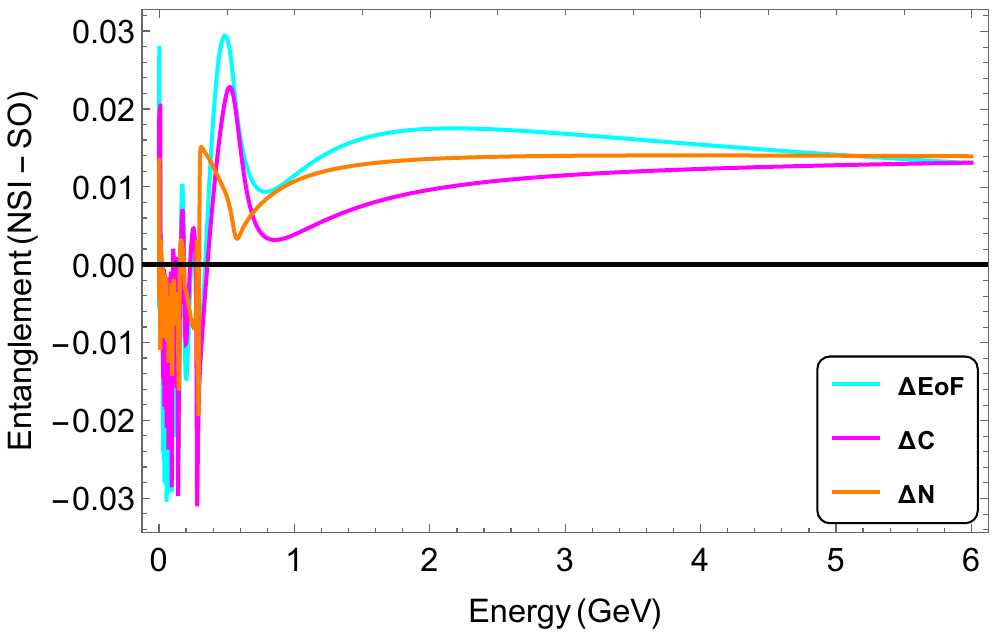}
\\
\includegraphics[scale=0.40]{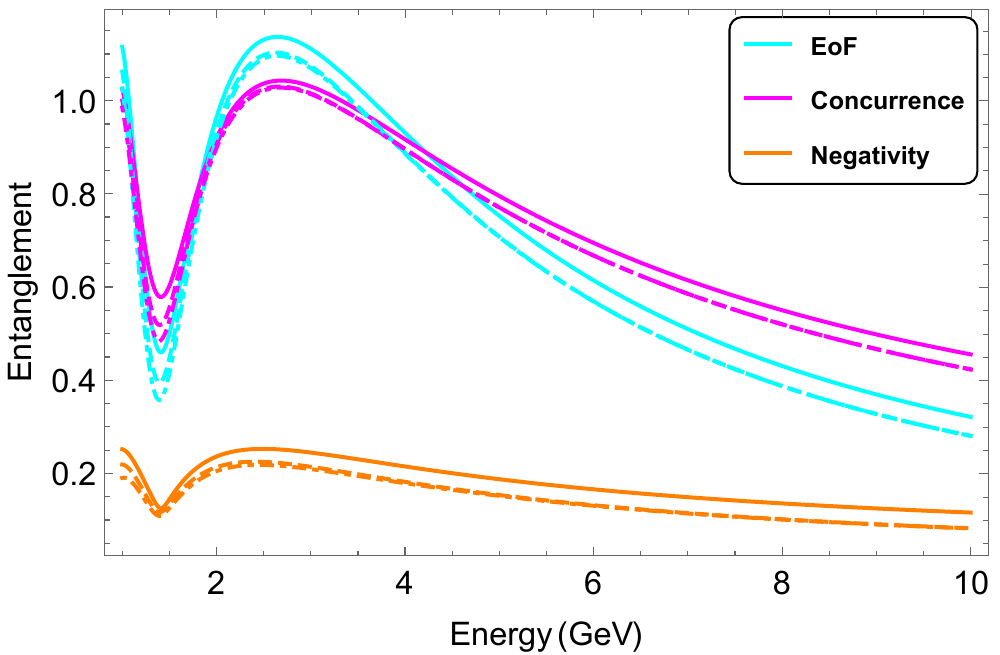}\hspace{1 cm}
\includegraphics[scale=0.40]{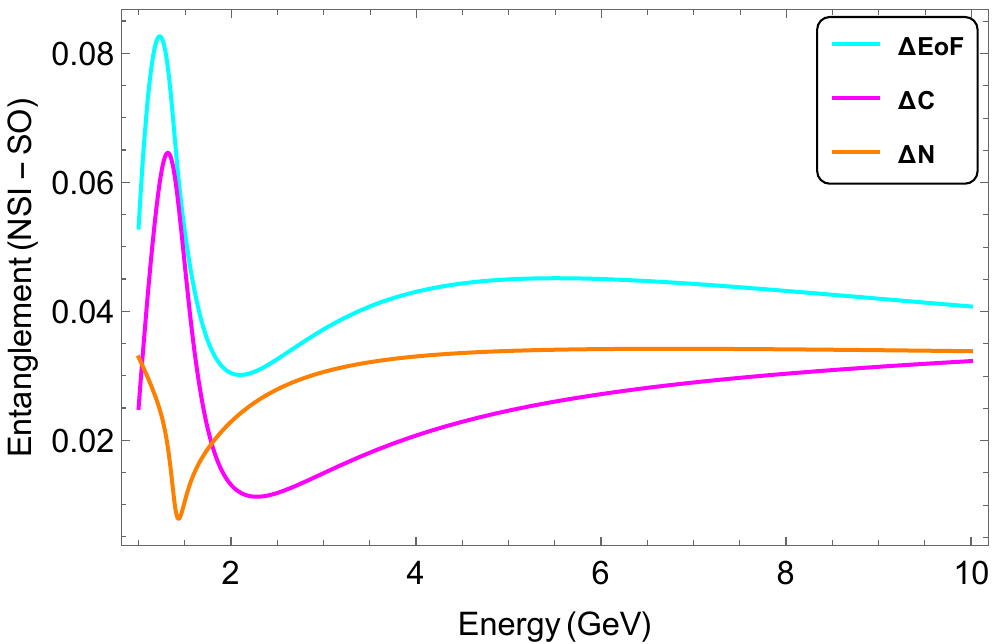}
\\
\includegraphics[scale=0.40]{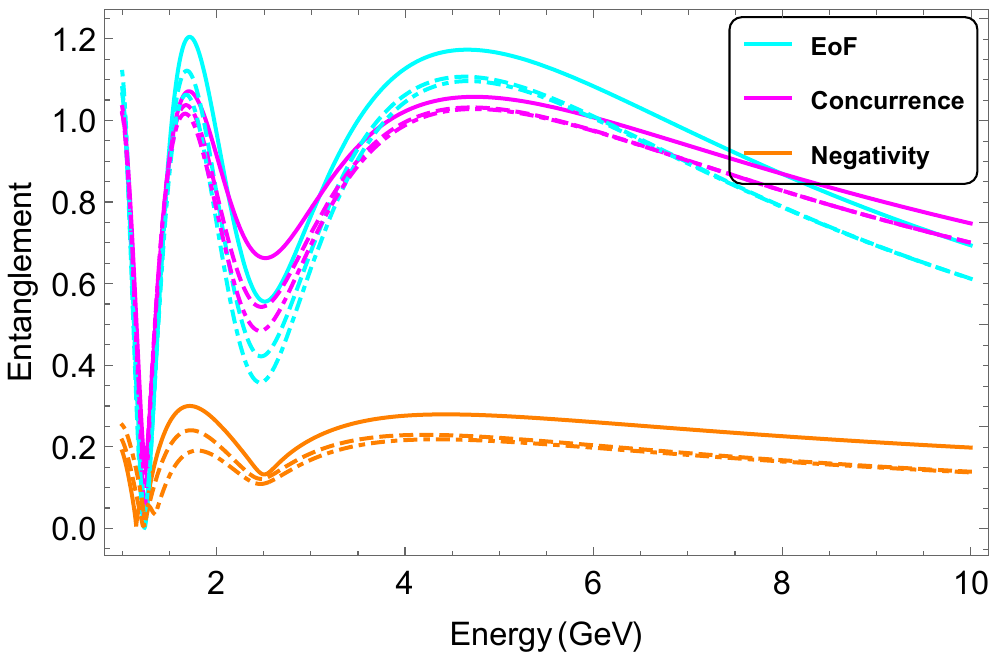}\hspace{1 cm}
\includegraphics[scale=0.40]{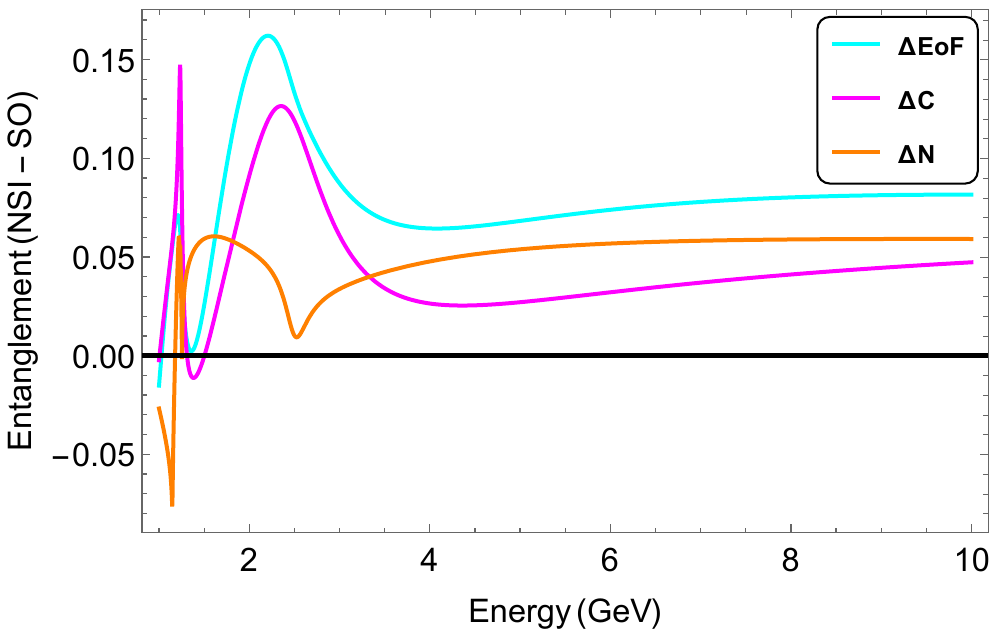}
\caption{The three panels depict the prediction of EOF, concurrence, and negativity for various accelerator neutrino oscillation experimental setups, i.e., T2K (upper), MINOS (middle), and DUNE (lower). Oscillation and NSI parameters are the same as given in figure \ref{fig3}. Colour representation follows cyan for EoF, magenta for concurrence, and orange for negativity.  In the left panel, NSI is represented by solid lines, the SO matter effect is represented by dashed lines, and the vacuum is represented by dot-dashed lines. }
\label{fig2}
\end{figure*}

In the right panel of Figure \ref{fig1}, it becomes evident that among the three entanglement measures, the negativity measure is the least influenced by the presence of NSI across all three reactor experimental configurations under consideration. Additionally, it is noteworthy that the impact of NSI is most subdued for the Daya Bay setup, whereas it is most pronounced for the KamLAND setup. This consistent trend holds true for all three entanglement measures. This observation can be attributed to the fact that the baseline distance is the greatest for the KamLAND configuration compared to the other two setups. The extended baseline enhances the sensitivity of the KamLAND setup to NSI effects, leading to a more pronounced influence on the entanglement measure. Overall, the effect of NSI is almost negligible in the reactor experiments. This result is also depicted from the left panel of Figure \ref{fig1} where the SO (dashed line) and NSI (solid line) curves are the same, while in the right panel, we can observe that the NSI effect is almost negligible.

 Figure \ref{fig2} illustrates the analysis of the three entanglement metrics within accelerator experiments, specifically T2K, MINOS, and DUNE.  In the left panel of the figure, it becomes evident that in the low-energy range, both EOF and concurrence can be regarded as robust measures of entanglement across all three experimental configurations, as their values closely align. However, in the high-energy domain, for T2K and MINOS experiments, concurrence exhibits slightly greater strength in comparison to EOF. Conversely, for the DUNE setup, the values of these two measures remain nearly identical, even for higher neutrino energy values. Similar to the reactor experimental setups considered, negativity consistently appears as a weaker measure for all three accelerator setups across the entire energy spectrum considered. Remarkably, these findings hold true for both matter SO and NSI interactions.

The right panel of Figure \ref{fig2} provides a closer look at the impact of NSI, characterized by the difference between values in the presence of NSI and SO matter interactions. Notably, the NSI effects are least pronounced for the T2K setup among all three configurations,  while they are most prominent for the DUNE experiment. The MINOS experiment experiences moderate NSI effects. This discrepancy can be attributed to differences in the baseline length. 

In equation \ref{EOF}, when NSI enhances $P_{\mu e}$ and reduces $\log_{2}P_{\mu e}$, the overall effect is that the first three terms increase while the last three terms decrease, relative to the SO matter effect. The primary impact arises from the first three terms in equation \ref{EOF}, leading to an increase in EoF for NSI. In equation \ref{C}, the term $3(P_{\alpha e}^{2}+P_{\alpha \mu}^{2}+P_{\alpha \tau}^{2})$ dominates and possesses a higher value for the SO matter effect compared to NSI, while the other terms have higher values for NSI than the matter effect. Conversely, in equation \ref{N}, the dependency of negativity on $P_{\mu e}$ outweighs that on the other two probabilities, and the impact of other terms is less significant. Hence, the enhancement for NSI is more pronounced in negativity. At lower energy ranges, $P_{\mu e}$ and $\sqrt{P_{\mu e}+P_{\mu \mu}}$ increase while other factors decrease and exert greater dominance. Consequently, the overall effect of NSI is a decrease in this scenario.

At E $\approx 3$ GeV, i.e., the energy corresponding to the maximum neutrino flux at DUNE, the overall strength of entanglement measures decreases while the impact of NSI is maximum on EOF and concurrence. Similarly, in MINOS, the energy at which the neutrino flux reaches its maximum $\approx 3$ GeV. At this energy, the strength of all three entanglement measures is maximum, and the NSI effect is moderate on all three measures. While in T2K, the energy value at which the neutrino flux is maximized is $\approx 0.6$ GeV, a similar feature like DUNE appears here; the strength of entanglement measures decreases at this value, but the effect of NSI on measures is maximum here. 

Overall, the effect of NSI is negligible in the T2K experiment as this is governed by oscillation in vacuum. This result is also depicted from the left panel of Figure \ref{fig2} where the SO (dashed line) and NSI (solid line) curve are almost the same, while in the right panel, we can observe that the NSI effect is very less as compared to other accelerator experiments.

In contrast to the three considered reactor experimental setups, it's observed that for all three accelerator experiments, NSI effects have the least impact on concurrence compared to EOF and negativity, particularly in the moderate and high-energy range of neutrinos. Specifically, EOF exhibits the highest sensitivity to NSI for the MINOS and DUNE experimental setups, whereas, for T2K, the NSI effects remain nearly the same for both EOF and negativity. In the low-energy region, the trends align with those observed in the reactor experiments, indicating that NSI effects are more pronounced for the EOF and concurrence measures as compared to negativity.

DUNE is expected to refine the acceptable values of NSI parameters \cite{kelly, bakhti, masud, dev, santiago}, thereby reducing the differences in the quantum correlations between the matter SO and NSI scenarios. Recently expected sensitivity on NC-NSI by DUNE is studied in \cite{santiago}. We also calculated the effect of these NC-NSI parameters in DUNE for a conservative scenario on quantum correlation measures. After DUNE sensitivity is taken into account, the value of EOF and concurrence reduces by approximately 60\%,  while the value of negativity reduces by approximately 40\% at peak values. For higher energy ranges, the value of EOF and negativity reduces by approximately 60\%,  while the value of concurrence reduces by approximately 55\%. 

The primary motive of this work was to study the impact of NSI on various entanglement measures. For this, we have obtained theoretical predictions of entanglement measures using the current bounds on NSI parameters. Based on these predictions, we have identified the maximum possible deviations in entanglement measures, which are allowed by the current data. However,  as these measures are expressed in terms of probabilities, such as $P_{\mu e}$, $P_{\mu \mu}$ and $P_{\mu \tau}$, the 
experimental measurement of these measures will be possible only if these probabilities are measured accurately.


\section{Conclusion}\label{sec5}
The impact of NSI on entanglement measures (EoF, concurrence, and negativity) is explored for three-flavor neutrino oscillation scenarios investigated in this study. Analysis is conducted across six neutrino experiments: T2K, MINOS, DUNE, Daya Bay, JUNO, and KamLAND. Among these, DUNE, with the longest baseline, exhibits a more pronounced deviation towards NSI compared to MINOS and T2K. Reactor experiments display lower sensitivity towards NSI, with negativity exhibiting the least sensitivity among entanglement measures. In contrast, accelerator experiments demonstrate greater sensitivity to NSI. Despite entanglement of formation and concurrence being effective measures in accelerator experiments, negativity still exhibits higher sensitivity than concurrence towards NSI.

\bibliographystyle{apsrev4-2}

\begin{thebibliography}{}

\bibitem{DUNE:2015lol}
R.~Acciarri \textit{et al.} [DUNE],
[arXiv:1512.06148 [physics.ins-det]].

\bibitem{T2K:2011qtm}
K.~Abe \textit{et al.} [T2K],
Nucl. Instrum. Meth. A \textbf{659}, 106-135 (2011)
[arXiv:1106.1238 [physics.ins-det]].

\bibitem{JUNO:2015zny}
F.~An \textit{et al.} [JUNO],
J. Phys. G \textbf{43}, no.3, 030401 (2016)
[arXiv:1507.05613 [physics.ins-det]].


\bibitem{peres1993}
C.~H.~Bennett, G.~Brassard, C.~Crépeau, R.~Jozsa, A.~Peres, and W.~K.~Wootters,
Phys. Rev. Lett, \textbf{70}, 1895 (1993).

\bibitem{ekert}
A.~K.~Ekert,
Phys. Rev. Lett. \textbf{67}, 661-663 (1991). 

\bibitem{Bennett}
C.~H.~Bennett, D.~P.~DiVincenzo, J.~A.~Smolin and  W.~K.~Wootters,
Phys. Rev. A \textbf{54}, no. 5, 3824 (1996).

\bibitem{Guo2020}
Y.~Guo, and L.~Zhang, 
Phys. Rev. A \textbf{101}, no. 3, 032301 (2020).



\bibitem{Wootters1998}
W.~K.~Wootters,  
Phys.  Rev. Lett. \textbf{80}, no. 10, 2245 (1998). 

\bibitem{Guo2019}
Y.~Guo, and G.~Gour,  
Phys. Rev. A \textbf{99}, no. 4, 042305 (2019).

\bibitem{Sabin2008}
C.~Sabín, and G.~García-Alcaine,   
Eur. Phys. J. C \textbf{48}, no. 3, 435-442 (2008).

\bibitem{Vidal2002}
G.~Vidal and R.~F.~Werner,  
Phys. Rev. A \textbf{65}, no. 3, 032314 (2002).



\bibitem{Blasone:2007wp}
M.~Blasone, F.~Dell'Anno, S.~De Siena, M.~Di Mauro and F.~Illuminati,
Phys. Rev. D \textbf{77}, 096002 (2008)
[arXiv:0711.2268 [quant-ph]].

\bibitem{Blasone:2007vw}
M.~Blasone, F.~Dell'Anno, S.~De Siena and F.~Illuminati,
EPL \textbf{85}, 50002 (2009)
[arXiv:0707.4476 [hep-ph]].

\bibitem{Banerjee:2014vga}
S.~Banerjee, A.~K.~Alok and R.~MacKenzie,
Eur. Phys. J. Plus \textbf{131}, no. 5, 129 (2016)
[arXiv:1409.1034 [hep-ph]].

\bibitem{Alok:2014gya}
A.~K.~Alok, S.~Banerjee and S.~U.~Sankar,
Nucl. Phys. B \textbf{909}, 65-72 (2016)
[arXiv:1411.5536 [hep-ph]].

\bibitem{Banerjee:2015mha}
S.~Banerjee, A.~K.~Alok, R.~Srikanth and B.~C.~Hiesmayr,
Eur. Phys. J. C \textbf{75}, no. 10, 487 (2015)
[arXiv:1508.03480 [hep-ph]].

\bibitem{Formaggio_2016}
J.~A.~Formaggio, D.~I.~Kaiser, M.~M.~Murskyj and T.~E.~Weiss, 
Phys.\ Rev.\ Lett. \textbf{117} no. 5, 050402(2016).

\bibitem{Fu:2017hky}
Q.~Fu and X.~Chen,
Eur. Phys. J. C \textbf{77}, no. 11, 775 (2017)
[arXiv:1705.08601 [hep-ph]].

\bibitem{Naikoo:2017fos}
J.~Naikoo, A.~K.~Alok, S.~Banerjee, S.~Uma Sankar, G.~Guarnieri, C.~Schultze and B.~C.~Hiesmayr,
Nucl. Phys. B \textbf{951}, 114872 (2020)
[arXiv:1710.05562 [hep-ph]].

\bibitem{Naikoo:2018vug}
J.~Naikoo, A.~K.~Alok and S.~Banerjee,
Phys. Rev. D \textbf{97}, no. 5, 053008 (2018)
[arXiv:1802.04265 [hep-ph]].

\bibitem{Naikoo:2019eec}
J.~Naikoo, A.~Kumar Alok, S.~Banerjee and S.~Uma Sankar,
Phys. Rev. D \textbf{99}, no. 9, 095001 (2019)
[arXiv:1901.10859 [hep-ph]].

\bibitem{Dixit:2019swl}
K.~Dixit and A.~Kumar Alok,
Eur. Phys. J. Plus \textbf{136}, no. 3, 334 (2021)
[arXiv:1909.04887 [hep-ph]].

\bibitem{Shafaq:2020sqo}
S.~Shafaq and P.~Mehta,
J. Phys. G \textbf{48}, no. 8, 085002 (2021)
[arXiv:2009.12328 [hep-ph]].


\bibitem{Ming:2020nyc}
F.~Ming, X.~K.~Song, J.~Ling, L.~Ye and D.~Wang,
Eur. Phys. J. C \textbf{80}, no. 3, 275 (2020)

\bibitem{Blasone:2021mbc}
M.~Blasone, F.~Illuminati, L.~Petruzziello and L.~Smaldone,
Phys. Rev. A \textbf{108}, no. 3, 032210 (2023)
[arXiv:2111.09979 [quant-ph]].

\bibitem{Yadav:2022grk}
B.~Yadav, T.~Sarkar, K.~Dixit and A.~K.~Alok,
Eur. Phys. J. C \textbf{82}, 446 (2022)
[arXiv:2201.05580 [hep-ph]].

\bibitem{Blasone:2022iwf}
M.~Blasone, F.~Illuminati, L.~Petruzziello, K.~Simonov and L.~Smaldone,
Eur. Phys. J. C \textbf{83}, no. 8, 688 (2023)
[arXiv:2211.16931 [hep-th]].

\bibitem{Chattopadhyay:2023xwr}
D.~S.~Chattopadhyay and A.~Dighe,
Phys. Rev. D \textbf{108}, no. 11, 112013 (2023)
[arXiv:2304.02475 [hep-ph]].

\bibitem{Blasone:2023gau}
M.~Blasone, S.~De Siena and C.~Matrella,
[arXiv:2305.06095 [quant-ph]].

\bibitem{Caban:2007je}
P.~Caban, J.~Rembielinski, K.~A.~Smolinski and Z.~Walczak,
Phys. Lett. A \textbf{363}, 389-391 (2007).

\bibitem{Alok:2015iua}
A.~K.~Alok, S.~Banerjee and S.~Uma Sankar,
Phys. Lett. B \textbf{749}, 94-97 (2015)
[arXiv:1504.02893 [hep-ph]].

\bibitem{Alok:2024amd}
A.~K.~Alok, S.~Banerjee, N.~R.~S.~Chundawat and S.~U.~Sankar,
[arXiv:2402.02470 [hep-ph]].

\bibitem{Capolupo:2018hrp}
A.~Capolupo, S.~M.~Giampaolo and G.~Lambiase,
Phys. Lett. B \textbf{792}, 298-303 (2019)
[arXiv:1807.07823 [hep-ph]].

\bibitem{Konwar:2024nwc}
L.~Konwar, J.~Vardani and B.~Yadav,
[arXiv:2401.02886 [hep-ph]].


\bibitem{DayaBay:2012fng}
F.~P.~An \textit{et al.} [Daya Bay],
Phys. Rev. Lett. \textbf{108}, 171803 (2012)
[arXiv:1203.1669 [hep-ex]].


\bibitem{Roskovec:2020rgr}
B.~Roskovec [Daya Bay],
PoS \textbf{ICHEP2020}, 170 (2021)



\bibitem{JUNO:2021vlw}
A.~Abusleme \textit{et al.} [JUNO],Prog. Part. Nucl. Phys., 10392,(2021)
[arXiv:2104.02565 [hep-ex]].


\bibitem{KamLAND:2002uet}
K.~Eguchi \textit{et al.} [KamLAND],
Phys. Rev. Lett. \textbf{90}, 021802 (2003)
[arXiv:hep-ex/0212021 [hep-ex]].




\bibitem{T2K:2013bzi}
K.~Abe \textit{et al.} [T2K],
Phys. Rev. Lett. \textbf{111}, no. 21, 211803 (2013)
[arXiv:1308.0465 [hep-ex]].


\bibitem{MINOS:2006foh}
D.~G.~Michael \textit{et al.} [MINOS],
Phys. Rev. Lett. \textbf{97}, 191801 (2006)
[arXiv:hep-ex/0607088 [hep-ex]].



\bibitem{DUNE:2020jqi}
B.~Abi \textit{et al.} [DUNE],
Eur. Phys. J. C \textbf{80}, no. 10, 978 (2020)
[arXiv:2006.16043 [hep-ex]].




\bibitem{Giunti:2007ry}
C.~Giunti and C.~W.~Kim, Oxford university press (2007).





\bibitem{Esteban:2018ppq}
I.~Esteban, M.~C~.Gonzalez-Garcia, M.~Maltoni, I.~Martinez-Soler and  J.~Salvado,
J. High Energy Phys., no. 8, 1-33 (2018).

\bibitem{Esteban:2019lfo}
I.~Esteban, M.~C~.Gonzalez-Garcia and M.~Maltoni,
J. High Energy Phys., no. 6, 1-24 (2019).

\bibitem{Coloma:2019mbs}
P.~Coloma, I.~Esteban, M.~C~.Gonzalez-Garcia and M.~Maltoni,
J. High Energy Phys., no. 2, 1-30 (2020).

\bibitem{Coloma2023}
P.~Coloma, M.~C~.Gonzalez-Garcia, M.~Maltoni, J.~P.~Pinheiro and S.~Urrea,  
arXiv preprint arXiv:2305.07698 (2023).

\bibitem{Ohlsson:1999xb}
T.~Ohlsson and H.~Snellman,
J. Math. Phys. \textbf{41}, 2768-2788 (2000)
[erratum: J. Math. Phys. \textbf{42}, 2345 (2001)]
[arXiv:hep-ph/9910546 [hep-ph]].


\bibitem{Jha}
A.~K.~Jha, S.~ Mukherjee and B.~A.~Bambah,  
Mod. Phys. Lett. A \textbf{36}, no. 09, 2150056 (2021).

\bibitem{jha2}
A.~K.~Jha, A.~Chatla and B.~A.~Bambah,  
Eur. Phys. J. Plus \textbf{139}, no. 1, 1-16 (2024).



\bibitem{Ming2021}
L.~J.~Li, F.~Ming, X.~ K.~Song, L.~Ye, and D.~Wang, 
Eur. Phys. J. C \textbf{81}, no. 8, 1-10 (2021).


\bibitem{Salas}
P.~F.~de Salas, D.~V.~Forero, S.~Gariazzo, P.~Mart\'\i{}nez-Mirav\'e, O.~Mena, C.~A.~Ternes, M.~T\'ortola and J.~W.~F.~Valle,
JHEP \textbf{02}, 071 (2021)
[arXiv:2006.11237 [hep-ph]].


\bibitem{kelly}
A.~de~Gouvêa and K.~J.~Kelly,  
Nucl. Phys. B, 908, 318-335 (2016).

\bibitem{bakhti}
P.~Bakhti, A.~N.~Khan and W.~Wang,   
J. Phys. G Nucl. Part. Phys. \textbf{44}, no. 12, 125001 (2017).

\bibitem{masud}
M.~Masud, S.~Roy, S. and P.~Mehta,  
Phys. Rev. D \textbf{99}, no. 11, 115032 (2019).

\bibitem{dev}
S.~S.~Chatterjee, P.~S.~Dev and P.~A.~Machado,   
J. High Energy Phys., no. 8, 1-25 (2021).

\bibitem{santiago}
A.~Cherchiglia and J.~Santiago, J.  
J. High Energy Phys., no. 3, 1-38 (2024).



\end{thebibliography}

\title{References}


\end{document}